\documentclass[aps,pra,twocolumn,superscriptaddress,nofootinbib,longbibliography,10pt]{revtex4-2}

\usepackage{graphicx}
\usepackage{amsfonts}
\usepackage{amssymb}
\usepackage{amsmath}
\usepackage{bm}
\usepackage{hyperref} 
\usepackage[dvipsnames]{xcolor}

\begin{document}

\title{Dissipative Quantum Vortices and Superradiant Scattering}%

\author{Thiago A. S. Cardoso}
\email{thiago.dasilvacardoso@ucdconnect.ie}
\affiliation{School of Mathematics and Statistics, University College Dublin, Belfield, Dublin 4, Ireland}%
\affiliation{Centro de Ci\^{e}ncias Naturais e Humanas, Universidade Federal do ABC (UFABC), 09210-580, Santo Andr\'{e}, S\~{a}o Paulo, Brazil}%
\author{Maur\'{i}cio Richartz}
 \email{mauricio.richartz@ufabc.edu.br}
\affiliation{Centro de Matem\'{a}tica, Computa\c{c}\~{a}o e Cogni\c{c}\~{a}o, Universidade Federal do ABC (UFABC), 09210-580, Santo Andr\'{e}, S\~{a}o Paulo, Brazil
}%

\begin{abstract}
Inspired by Analogue Gravity, superradiance has been previously investigated in Bose-Einstein condensates (BECs).
In this work, we revisit this problem by introducing dissipation to the system. After establishing the possibility of quantum vortices in dissipative BECs, we analyze the propagation of elementary excitations and demonstrate the existence of superradiant modes which can be interpreted in terms of the dissipation of ``antiparticles".
Our findings support the possibility of superradiant scattering around dissipative quantum vortices and paves the way for future experimental realization of the phenomenon.
\end{abstract}

\maketitle

\section{\label{intr}Introduction}
Rotational superradiance~\cite{superreview} is the phenomenon in which low-frequency waves impinging on a rotating scattering object are amplified upon reflection. The process is associated with an outgoing flux of energy and angular momentum from the scatterer, a fact which translates into the reflected waves having greater amplitudes than the incident ones. It typically occurs when the wave frequency $\omega$, the azimuthal wave number $\lambda$ and the angular velocity $\Omega$ of the object satisfy the inequality
\begin{equation}
\omega - \lambda \Omega < 0.
\end{equation}
Two canonical situations in which superradiance is possible are the scattering of electromagnetic radiation by a metal cylinder (referred to as Zel'dovich's cylinder)~\cite{zeldovich1,zeldovich2,bekenstein} and the scattering of bosonic fields by rotating black holes~\cite{misner,staro1,staro2}. 

The necessary ingredients for superradiance are the existence of negative energy states in the system and a mechanism to dissipate them~\cite{Richartz:2009mi}. Rotational motion introduces the possibility of negative energies since observers in the laboratory frame will measure lower energies than in the rotating frame. 
In the black hole scenario, dissipation is associated with the existence of the event horizon, which acts as a one-way membrane and can effectively eliminate negative energy modes from the system. In the case of Zel'dovich's cylinder, dissipation is a consequence of the Joule effect associated with the conductivity of the cylinder. If negative energy modes are present, but no mechanism to dissipate them exist, one expects instabilities to arise. In General Relativity, such instabilities are known as ergosphere instabilities~\cite{ergoI}.

New insights into the phenomenon of superradiance have been provided in the last two decades by Analogue Gravity~\cite{volovik1995there,reviewvisser,AGbook,Barcelo:2018ynq,Jacquet:2020bar}. The Analogue Gravity programme was initiated by Unruh~\cite{unruhI} and consists in using condensed matter systems to simulate aspects of gravitational physics, including black holes, in the laboratory. Although very important theoretical advances have been obtained, it is on the experimental side that the greatest successes of Analogue Gravity have been achieved~\cite{Weinfurtner:2010nu,Euve:2015vml,MunozdeNova:2018fxv,Drori:2018ivu,Kolobov:2019qfs,Torres_2017,2020NatPh..16.1069C,Braidotti:2021nhw,Torres:2020tzs,Eckel:2017uqx,Jaskula_2012,Prain:2017wlo,Steinhauer:2021fhb,Viermann:2022wgw,Barroso:2022vxg}. In particular, the first observation of rotational superradiance was performed by studying surface waves on a water tank~\cite{Torres_2017}, demonstrating the robustness of the effect -- see also \cite{2020NatPh..16.1069C,Braidotti:2021nhw}. 
Another milestone for Analogue Gravity was the detection of the Hawking effect in a Bose-Einstein condensate (BEC)~\cite{MunozdeNova:2018fxv,Isoard:2019buh} -- see also \cite{Weinfurtner:2010nu,Euve:2015vml,Drori:2018ivu,Kolobov:2019qfs}.

Previous work on BEC superradiance, under the Analogue Gravity framework, has been performed in the non-dispersive regime, i.e.~assuming the Thomas-Fermi (or hydrodynamical) approximation~\cite{superrad_bec1,superrad_bec2,superrad_bec22,superrad_bec3,superrad_bec4,superrad_bec5,superrad_bec6}. 
The occurrence of the ergoregion instability in BECs has also been investigated in the non-dispersive regime~\cite{ergo2,ergo3,ergo4}. The main limitation of this approach, however, is that one needs to ignore quantum effects in order to establish the analogy. More recently, going beyond Analogue Gravity and taking into account the full Bogolyubov dispersive regime, Refs.~\cite{Richartz:2012bd,carusottoI,sam_dispersive} reanalyzed the possibility of superradiance in BECs. In particular, Ref.~\cite{carusottoI} determined that the existence of negative energy states, a natural feature of elementary excitations in BECs, is associated to instabilities of the ``ergoregion type", analogous to the ones from General Relativity~\cite{ergoI}. Superradiant modes were then studied analytically and numerically in a non-rotating configuration~\cite{carusottoII,carusottoIII}.

Inspired by theoretical models of an atom laser~\cite{atomlaserI,atomlaserII,sebastianIV}, we consider the possibility of Zeldovich's superradiance in dissipative quantum vortices. In atom lasers the goal is to create an intense coherent beam of atoms (analogous to the coherent beam of light in photon lasers). As the atoms in a BEC already exhibit coherent behavior, they can be used to create the beam. A coupling between the condensed phase and the excited states is provided by manipulating the trap holding the system in place and in this way particles are taken out of the BEC. Such extraction of atoms corresponds to dissipation. In this work, taking into account the general framework for superradiance established in Ref.~\cite{Richartz:2012bd}, we introduce localized dissipation in a BEC as in Refs.~\cite{Brazhnyi_2009,Barontini_2013}. We then obtain the associated quantum vortex solutions and find that superradiant modes naturally arise from the usual BEC second-quantization formalism. We emphasize a simple physical interpretation of these modes as ``particles" and ``antiparticles" and demonstrate that superradiance takes place when ``antiparticles" are dissipated. The existence of stationary superradiant modes in our analysis is independent of the notion of an (analogue) event horizon, as in the case of Zel'dovich's cylinder~\cite{zeldovich1,zeldovich2,bekenstein} and rotating stars~\cite{Richartz:2013unq,Cardoso:2015zqa,Cardoso:2017kgn,Day:2019bbh,Chadha-Day:2022inf}.

The present article is organized as follows. In Sec.~II, we set the notation  and the framework for studying BECs and their elementary excitations in the presence of dissipation. In particular, we discuss under which conditions dissipation can give rise to superradiance. In Sec.~III we specify a particular dissipation profile and show that the associated condensate is described by a quantized vortex. In Sec.~IV we analyze (theoretically and numerically) the scattering of perturbations by the dissipative vortex and investigate the ocurrence of superradiance in the system. Sec.~V concludes the work with our final remarks.   

\section{\label{theBEC} BEC Theory with dissipation}
\subsection{\label{theBEC:GPE}Gross-Pitaevskii Equation}
When a system of identical bosons is cooled down to sufficiently low temperatures, it undergoes a phase transition in which a macroscopic number of particles occupies the ground state of the system. Under the second quantization formalism, let $\hat{\psi}(t,\bm{r})$ and $\hat{\psi}^\dagger(t,\bm{r})$ denote, respectively, the bosonic field operator and its Hermitian conjugate. A physical system of $N$ identical bosons of mass $m$ can then be described by the following Hamiltonian operator~\cite{fw,BECbook,pathria}:
\begin{align}\label{hamboson}
\nonumber\hat{H}&=\int\hat{\psi}^\dagger(t,\bm{r})\left[-\frac{\hbar^2}{2m}\bm{\nabla}^2+V_{ext}(\bm{r}) + i\frac{\hbar}{2} \Gamma(r)\right]\hat{\psi}(t,\bm{r})\text{ }d\bm{r}\\
&+\frac{U_0}{2}\int\hat{\psi}^\dagger(t,\bm{r})\hat{\psi}^\dagger(t,\bm{r})\hat{\psi}(t,\bm{r})\hat{\psi}(t,\bm{r})\text{ }d\bm{r},
\end{align}
where $V_{ext}(\bm{r})$ is the external confinement potential, $U_0$ is associated to the strength of the interparticle interaction~\cite{BECbook,pathria} and $\hbar$ is the reduced Planck's constant. The quantity $\Gamma(\bm{r})$ measures the rate at which bosons are fed into ($\Gamma>0$) or taken out of ($\Gamma<0$) the system~\cite{sebastianI,garay2,sebastianIII,sebastianIV}. Physically, it models the interaction of the BEC itself with other systems in contact with it, like the non-condensed phase or the magnetic trap holding it in place \cite{garay2,sebastianIII}. It is also the starting point for studying atom lasers \cite{atomlaserI,atomlaserII}.

The operator $\hat{H}$ describes a dilute gas, since the interaction integral only involves pairs of bosons.
In the Heisenberg representation, the field operator evolves in time according to
\begin{equation}\label{heis}
\frac{d\hat{\psi}}{dt}=\frac{1}{i\hbar}\left[\hat{\psi},\hat{H}\right] = \frac{1}{i\hbar} \left(\hat{\psi}\hat{H}-\hat{H}\hat{\psi}\right).
\end{equation}
However, since only states around the condensed phase will be important in the following, it is convenient to take Bogolyubov's approximation \cite{BECbook}:
\begin{equation}\label{bogoapprox}
\hat{\psi}(t,\bm{r})\approx\psi_0(t,\bm{r})\mathbb{I}+\hat{\psi_1}(t,\bm{r}),
\end{equation}
where $\mathbb{I}$ is the identity operator, $\psi_0(t,\bm{r})$ is the mean field representing the condensate, and $\hat{\psi_1}(t,\bm{r})$ accounts for fluctuations of the condensate. In this approximation, the number of bosons occupying the condensed phase is large enough that a small depletion and/or accretion of particles effectively does not change the condensed phase. In other words, the ground state contribution to the field operator is approximated by the complex function $\psi_0(t,\bm{r})$, commonly known as the wave function of the condensate.

The equation of motion for the mean field, obtained by substituting Eq.~\eqref{bogoapprox} into Eq.~\eqref{heis} and keeping only zeroth order terms, is the Gross-Pitaevskii equation (as considered in Refs.~\cite{Brazhnyi_2009,Barontini_2013}):
\begin{align}\label{GPEloss}
i\hbar\frac{\partial\psi_0}{\partial t}=-\frac{\hbar^2}{2m}\bm{\nabla}^2\psi_0+&\left[V_{ext}(\bm{r})+i\frac{\hbar}{2}\Gamma(\bm{r})\right]\psi_0\notag\\
+&U_0|\psi_0|^2\psi_0.
\end{align}
It consists of a nonlinear Schr\"{o}dinger equation describing the dynamics of the condensed phase in a BEC.
Although the wave function of the condensate is a complex number, it is related to real observables of the system. This becomes evident in Madelung's representation $\psi_0(t,\bm{r})=\sqrt{\rho_0(t,\bm{r})}e^{iS_0(t,\bm{r})}$, which transforms Eq.~\eqref{GPEloss} into
\begin{equation}\label{dissMadGPEI}
\frac{\partial \rho _0}{\partial t}+\bm{\nabla}\cdot\left(\rho _0 \bm{v}_0 \right)=\rho_0\Gamma, 
\end{equation}
and
\begin{equation}\label{dissMadGPEII}
\frac{\partial S_0 }{\partial t} +\frac{1}{2}|\bm{ {v}_0}|^2 + P_0+\frac{V_{ext}(\bm{r})}{m}=0,
\end{equation}
where $\bm{v}_0 = (\hbar/m)\bm{\nabla}S_0$ and
\begin{equation}
 P_0 = \frac{U_0}{m}\rho_0-\frac{\hbar^2}{2m^2}\frac{1}{\sqrt{\rho_0}}\bm{\nabla}^2\sqrt{\rho_0}.
\end{equation}

We highlight that Eq.~\eqref{dissMadGPEI} is a general continuity equation that has a source (sink) term proportional to the gain (loss) rate $\Gamma$. In the absence of dissipation ($\Gamma = 0$), Eqs.~\eqref{dissMadGPEI} and \eqref{dissMadGPEII} are, respectively, the continuity equation and the Bernoulli equation for an ideal fluid if we interpret $\rho_0$ as the density of particles, $\bm{v}_0$ as the velocity field of the condensate, and $P_0$ as the pressure of the condensate. Thus, a BEC is a fluid in which quantum properties are manifest on a macroscopic scale. In particular, the condensate flow is irrotational ($\bm{\nabla}\times\bm{v}=0$) and has no viscosity. The term proportional to $\hbar^2$ in $P_0$ is known as the quantum pressure, and violates barotropicity. When the quantum pressure is negligible, velocity perturbations of the condensate obey the equation of motion for a real scalar field in a curved spacetime~\cite{garay1}, thus motivating the study of BECs as analogue models of gravity.

\subsection{\label{Exc}Elementary Excitations}

Excitations of a BEC can by investigated through the fluctuation operator $\hat{\psi_1}(t,\bm{r})$ in the Bogolyubov approximation \eqref{bogoapprox}. We adopt the same notation as in Refs.~\cite{leonhardt,carusottoI} by defining a new operator $\hat{\delta\psi}(t,\bm{r})$, in terms of $\hat{\psi_1}(t,\bm{r})$, as 
\begin{equation}
\hat{\delta\psi}(t,\bm{r}) = e^{-iS_0(t,\bm{r})} \hat{\psi_1}(t,\bm{r}) .  
\end{equation}
The main difference of this section in comparison to the formalism presented in previous work~\cite{leonhardt,carusottoI} is the possibility of dissipation through the function $\Gamma$. It is straightforward to check that both $\hat{\delta\psi}(t,\bm{r})$ and its Hermitian conjugate 
will appear as first-order contributions when one substitutes Eq.~\eqref{bogoapprox} into Eq.~\eqref{heis}. It is convenient to treat them as one single spinor quantum field $\Psi$ such that:
\begin{equation}\label{qspin}
\Psi=\left[\begin{array}{c}
\displaystyle\hat{\delta\psi}\\
\text{ }\\
\displaystyle\hat{\delta\psi}^\dagger
\end{array}\right], \qquad \Psi^\dagger = \left[\begin{array}{c}
\displaystyle\hat{\delta\psi}^\dagger\\
\text{ }\\
\displaystyle\hat{\delta\psi}
\end{array}\right].
\end{equation}
The spinor $\Psi$ has the symmetry $\overline{\Psi}\equiv\sigma_x\Psi^\dagger=\Psi$, where $\sigma_x$ is one of the Pauli matrices given by
\begin{equation}
\sigma_x=\left[\begin{array}{cc}0 & 1\\1 & 0\end{array}\right], \quad \sigma_y=\left[\begin{array}{cc}0 & -i\\i & 0\end{array}\right], \quad \sigma_z=\left[\begin{array}{cc}1 & 0\\0 & -1\end{array}\right].
\end{equation}

An expansion into stationary states can be performed according to
\begin{equation}\label{expspin}
\Psi(t,\bm{r})=\displaystyle\sum_j\left(\hat{a}_jW_j(t,\bm{r})+\hat{a}_j^\dagger\overline{W}_j(t,\bm{r})\right),
\end{equation}
where $\hat{a}_j$ and $\hat{a}_j^\dagger$ are, respectively, bosonic annihilation and creation operators, $W_j(t,\bm{r})$ is a spinor, and $j$ indexes the modes. Being bosonic operators, $\hat{a}_i$ and $\hat{a}^\dagger_j $ satisfy the following commutation relations
\begin{equation}
[\hat{a}_i,\hat{a}_j]=0, \quad [\hat{a}^\dagger_i,\hat{a}^\dagger_j]=0, \quad [\hat{a}_i,\hat{a}^\dagger_j]=\delta_{ij}\mathbb{I},
\end{equation}
where $\delta_{ij}$ is the Kronecker delta. The spinors $W_j(t,\bm{r})$ and $\overline{W}_j(t,\bm{r})$ are written in terms of complex functions $u_j(t,\bm{r})$ and $v_j(t,\bm{r})$ as  
\begin{equation}
W_j(t,\bm{r})=\left[\begin{array}{c} u_j(t,\bm{r})\\ v_j(t,\bm{r})\end{array}\right], \quad \overline{W}_j(t,\bm{r})=\left[\begin{array}{c} v_j^*(t,\bm{r})\\ u_j^*(t,\bm{r})\end{array}\right],
\end{equation}
where $^*$ denotes complex conjugation.

Using expansion \eqref{expspin}, the first-order terms arising from Eq.~\eqref{heis} lead to the following equation for the perturbations:
\begin{equation}\label{pertspinI}
i\hbar\left(\frac{\partial}{\partial t}+\bm{v}_0\cdot\bm{\nabla}+\frac{1}{2}\bm{\nabla}\cdot\bm{v}_0\right)W_j=\left( \hat{\mathcal{H}}+i\frac{\hbar}{2}\Gamma \right)W_j,
\end{equation}
where 
\begin{align}\label{pertspinII}
\hat{\mathcal{H}}=\bigg[&-\frac{\hbar^2}{2m}\bm{\nabla}^2+\frac{m\bm{v}_0^2}{2}+V_{ext}(\bm{r})+ \hbar \frac{\partial S_0}{\partial t} +2U_0\rho_0(\bm{r})\bigg]\sigma_z\notag\\&+iU_0\rho_0(\bm{r})\sigma_y.
\end{align}
Note that there is a mixing of the components of the spinor through the last term involving $\sigma_y$.

The Hamiltonian-like operator $\hat{\mathcal{H}}$ is not Hermitian. However, it satisfies $\sigma_z\hat{\mathcal{H}}^\dagger\sigma_z=\hat{\mathcal{H}}$. We use this fact and define
\begin{equation}
W_j^\dagger(t,\bm{r})\equiv\left[\begin{array}{cc}\displaystyle \! u_j^*(t,\bm{r})&\displaystyle v_j^*(t,\bm{r})\end{array}\right]
\end{equation}
to show that Eq.~\eqref{pertspinI} implies the conservation equation
\begin{equation}\label{disscontspin}
\frac{\partial}{\partial t}\left(W_j^\dagger \sigma_z W_j \right)+\bm{\nabla}\cdot\bm{J}_j=\Gamma(\bm{r})W_j^\dagger\sigma_z W_j,
\end{equation}
where $\bm{J}_j$ is the current defined by
\begin{equation}\label{contspinII}
\bm{J}_j\equiv\bm{v}_0W_j^\dagger\sigma_zW_j+\frac{\hbar}{2mi}\left(W_j^\dagger\bm{\nabla}W_j-\bm{\nabla}W_j^\dagger W_j\right).
\end{equation}
The conservation equation above suggests the following inner product in the solution-space of Eq.~\eqref{pertspinI}:
\begin{equation}\label{ip}
\langle W_i,W_j \rangle \equiv\int\left(u_i^*u_j-v_i^*v_j\right)d\bm{r}.
\end{equation}
In fact, if boundary terms can be neglected, the (squared) norm of any solution, e.g.~$\langle W_j,W_j \rangle$, is conserved in time when $\Gamma = 0$.

We remark that the inner product \eqref{ip} is not a positive-definite quantity. Indeed, $u_j$ gives a positive contribution to it, while $v_j$ is responsible for a negative contribution. If expansion \eqref{expspin} is built up using stationary solutions $W_\omega$ labeled by their positive real frequencies $\omega$, it can be shown \cite{BECbook,leonhardt,carusottoI} that the energy of one such spinor excitation is given by
\begin{equation}\label{enexc}
E_\omega=\hbar\omega \langle W_\omega,W_\omega \rangle.
\end{equation}
 Thus, the energy of an excitation depends on the sign of its (squared) norm: modes with positive norm have positive energy, while modes with negative norm have negative energy. This fact points to a simple interpretation for the spinor excitations $W_\omega(t,\bm{r})$: $u_\omega(t,\bm{r})$ acts as a particle and $v_\omega(t,\bm{r})$ acts as an anti-particle. The quantity $W_\omega ^\dagger\sigma_zW_\omega$ can be interpreted as the ``charge" density, which is locally conserved according to Eq.~\eqref{disscontspin} when $\Gamma=0$.

It is evident from Eq.~\eqref{disscontspin} that, not only the background, but also the excitations are subject to losses when $\Gamma \neq 0$. An interesting feature of such losses is that the term $\Gamma W_j^\dagger\sigma_zW_j$ in Eq.~\eqref{disscontspin} can be positive even if $\Gamma$ is negative. 
More precisely, if the ``charge" density is locally negative, the perturbations experience an increase in energy even though atoms are being taken out of the system. This is closely associated to dissipation of negative energy states, which points to the existence of superradiance in the system. We explore this idea in Sec.~\ref{diss:super}, where we discuss the possibility of superradiant scattering for perturbations around the quantized vortex solution that we obtain in Sec.~\ref{theBEC:vortex}.

\section{\label{theBEC:vortex}Quantized Vortex}
A standard textbook solution of the Gross-Pitaevskii equation \eqref{GPEloss} without dissipation (i.e.~$\Gamma = 0$) is the quantized vortex~\cite{fw,BECbook,pathria}. In this section we analyze how dissipation affects the well-known vortex solution, thus obtaining configurations that describe dissipative quantum vortices. We assume that the BEC is two-dimensional and infinite, and set $V_{ext}(\bm{r})=0$. We also assume that the system is axisymmetric and stationary.

Adopting standard polar coordinates $\bm{r}=(s,\varphi)$ we introduce the following ansatz for the Madelung variables:
\begin{equation} \label{md1}
\rho_0(t,\bm{r}) = \rho_{\infty} \left[F(s)\right]^2 
\end{equation}
and
\begin{equation} \label{md2}
S_0(t,\bm{r}) = - \frac{\mu}{\hbar} t + \nu\varphi + f(s), 
\end{equation}
where $\rho_\infty$ is the density of the condensate at infinity, $\mu$ is the chemical potential, and $\nu$ is the azimuthal number (which must be an integer in order for the wave function to be single-valued). Without loss of generality we assume that $\nu \in \mathbb Z_+$. The functions $F(s)$ and $f(s)$ are auxilliary dimensionless functions. 
The associated velocity field is given by
\begin{equation}\label{disswfII}
    \bm{v}_0=v_s(s)\,\bm{e}_s+v_\varphi(s)\,\bm{e}_\varphi=\frac{\hbar}{m}\left[f'(s)\,\bm{e}_s+\frac{\nu}{s}\,\bm{e}_\varphi\right],
\end{equation}
where $\bm{e}_s$ and $\bm{e}_\varphi$ are the unit vectors in the $s$ and $\varphi$ directions, respectively. 

We remark that one can add a constant to $f(s)$ without affecting the background velocity. We eliminate this gauge freedom by assuming that $f(s) \rightarrow 0$ as $s \rightarrow \infty$. Note also that circulation is quantized in multiples of the fundamental flux unit $2\pi \hbar/m$, as in the standard case. Here, however, we also allow for a non-zero radial velocity since atoms can flow into or out of the condensate. In fact, inspired by Zel'dovich's cylinder~\cite{zeldovich1,zeldovich2}, we assume that the atoms are taken out of the system, and that this process takes place mainly at the innermost regions of the vortex. This translates into the dissipation parameter $\Gamma=\Gamma(s)$ being negative and concentrated around the origin ($s=0$).

\begin{figure}[!htbp]
\centering
  \includegraphics[width = 0.99 \linewidth]{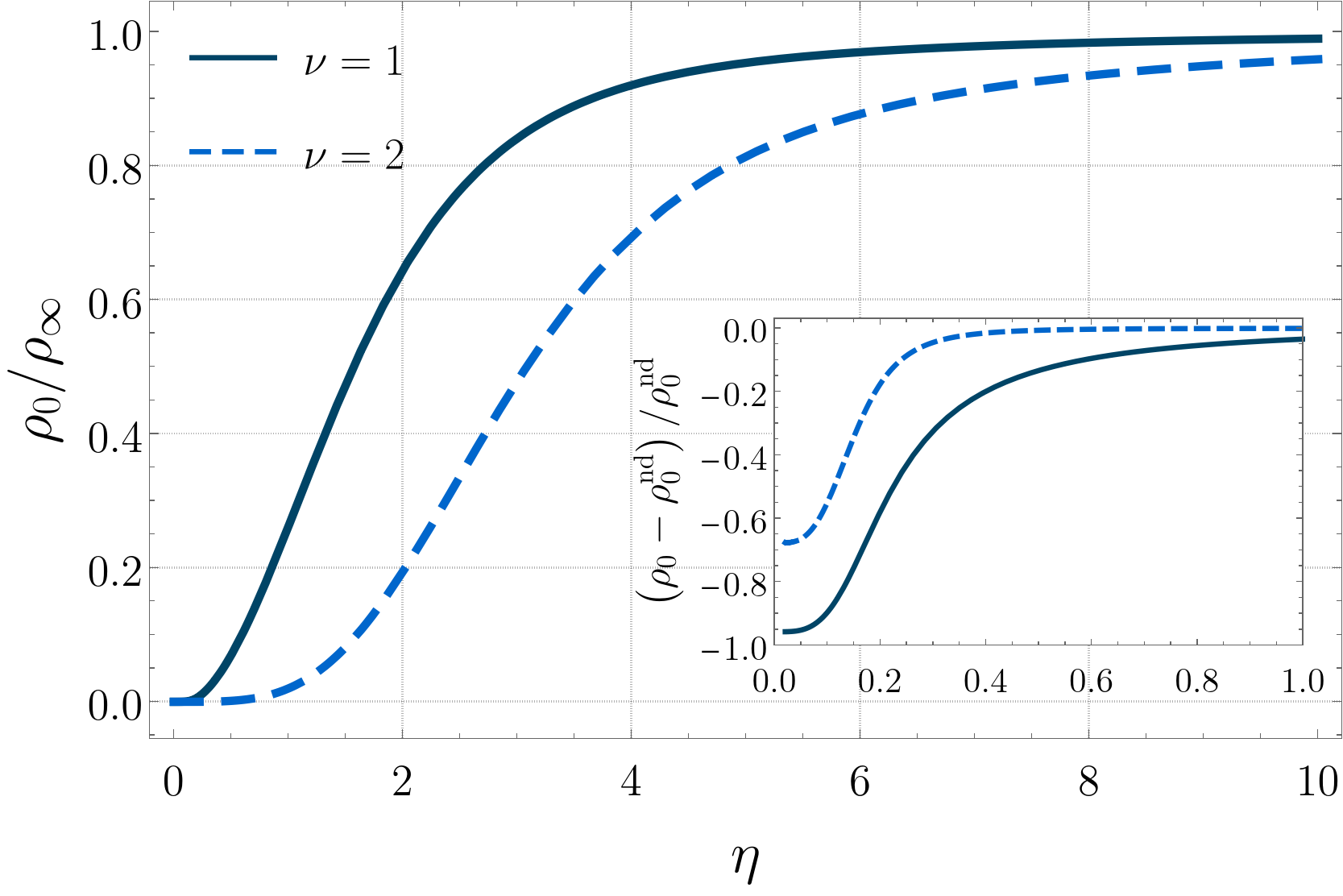}
  \caption{Normalized density profile $\rho_0/\rho_\infty$ of a quantized BEC vortex with dissipation for $\nu=1$ (solid curve) and $\nu=2$ (dashed curve). Dissipation is characterized by the Gaussian function \eqref{gaussd} with $P_1=0.1$ and $P_2=3000$. The inset shows the relative difference between the density $\rho_0$ of the dissipative vortex and the density $\rho_0^{\mathrm{nd}}$ of the corresponding standard non-dissipative vortex.}
  \label{dens_diss_final} 
\end{figure}
\begin{figure}[!htbp]
\centering
  \includegraphics[width = 0.99 \linewidth]{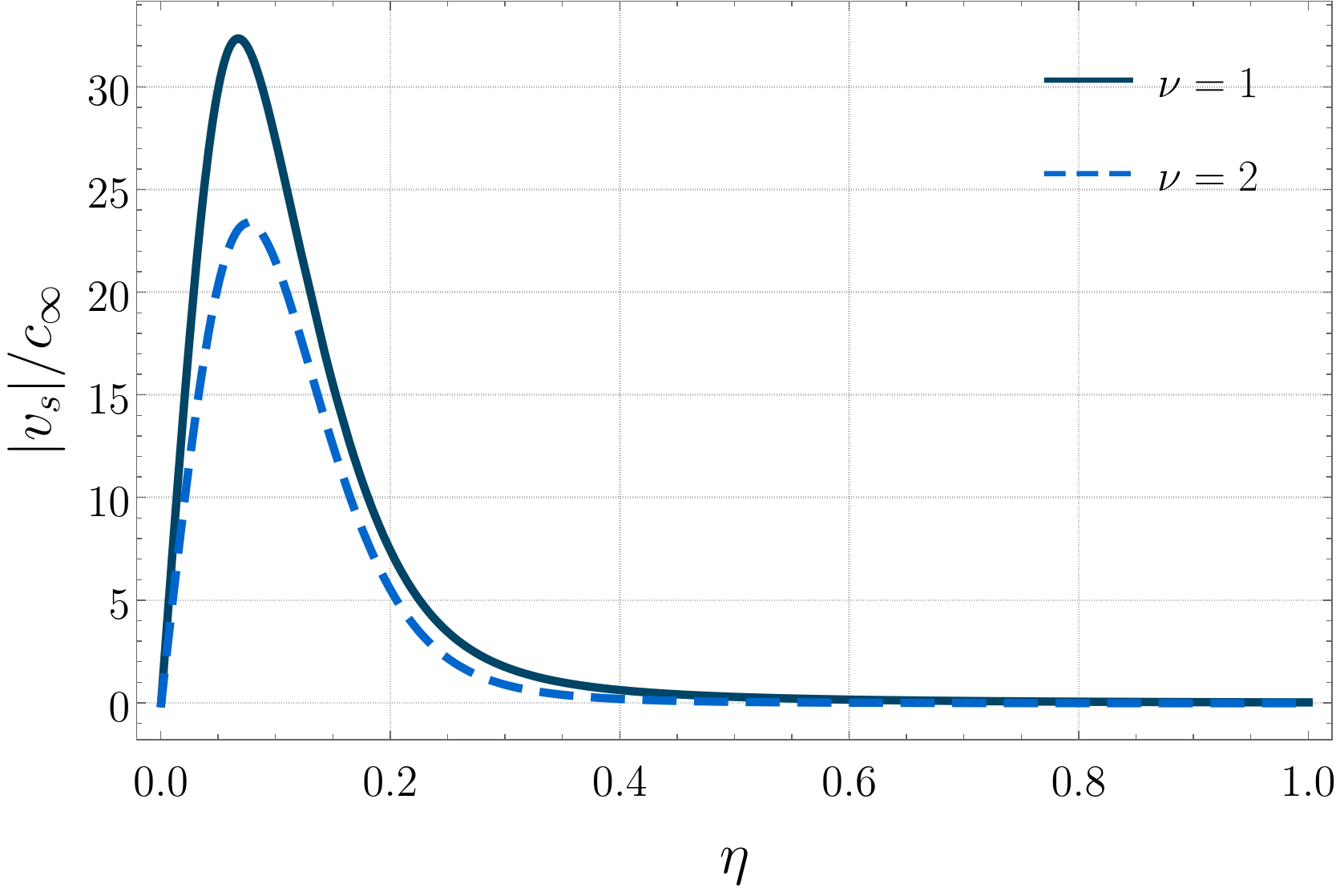}
  \caption{Normalized radial velocity $|v_s|/c_{\infty}$ of a quantized BEC vortex with dissipation for $\nu=1$ (solid curve) and $\nu=2$ (dashed curve). Dissipation is characterized by the Gaussian function \eqref{gaussd} with $P_1=0.1$ and $P_2=3000$.}
    \label{vel_diss_final}
\end{figure}

  At this point, it is convenient to define the dimensionless variable $\eta=s/\xi$, where $\xi\equiv\hbar/\sqrt{2mU_0\rho_\infty}$. The parameter $\xi$, known as the healing length, is a typical length parameter for BECs. The density and velocity profiles of the vortex can then be found by inserting Eqs.~\eqref{md1} and \eqref{disswfII} into Eqs.~\eqref{dissMadGPEI} and \eqref{dissMadGPEII}, yielding:
\begin{equation}\label{addissMaGPEI}
\frac{1}{\eta}\frac{d}{d\eta}\left(\eta F^2\widetilde{v}_s\right)=F^2\widetilde{\Gamma}
\end{equation}
and
\begin{equation}\label{addissMaGPEII}
\frac{1}{\eta}\frac{d}{d\eta}\left(\eta\frac{dF}{d\eta}\right)+\left(1-\frac{\nu^2}{\eta^2}\right)F-F^3-\frac{1}{2}F\widetilde{v}_s^2=0,
\end{equation}
where $\widetilde{\Gamma} = \xi \Gamma / c_\infty$ and $\widetilde{v}_s =  v_s/c_\infty$ are dimensionless quantities. The parameter $c_\infty$, defined by $c_\infty^2=U_0 \rho_\infty /m$, can be interpreted as the speed of the perturbations at infinity ($\eta \rightarrow \infty$).

By analyzing the behaviour of the differential equations \eqref{addissMaGPEI} and \eqref{addissMaGPEII} in the limit $\eta \rightarrow 0$, one can show that a well-behaved solution must satisfy the boundary condition
\begin{equation} \label{bcc1}
\widetilde v_s(0)=0.
\end{equation}
On the other hand, far away from the vortex, we assume that the density approaches a constant value and that the velocity vanishes. In our notation, this translates into
\begin{equation} \label{bcc2}
\lim\limits_{\eta \to\infty}F(\eta)=1, \qquad \lim\limits_{\eta \to\infty}\widetilde{v}_s(\eta)=0.
\end{equation}

Throughout this work we use experiment-inspired values for the physical parameters~\cite{BECreview}:
\begin{equation}
\begin{cases}
\rho_\infty=10^{13}\, \mathrm{cm}^{-3}, \\
 m=87\, \mathrm{a.u.}, \\
 U_0=2.21\times10^{-49}\, \mathrm{J.m}^2.
\end{cases}
 \end{equation}
In order to illustrate the effects of dissipation, we choose the dissipation profile to be a Gaussian function. More specifically, we set         
\begin{equation} \label{gaussd}
\widetilde{\Gamma}(\eta)=-P_2 \exp{\left(- \eta^2/P_1 ^2\right)},
\end{equation}
where $P_1$ and $P_2$ are dimensionless constants. The spread parameter $P_1=0.1$ and the height parameter $P_2=3000$ are chosen so that the flux of dissipated atoms has the same order of magnitude as in a real atom laser~\cite{ultrabright}. 
 
Eqs.~\eqref{addissMaGPEI} and \eqref{addissMaGPEII} are solved both in the dissipative and non-dissipative scenarios using the pseudo-spectral method described in the Appendix. In Fig.~\ref{dens_diss_final} we plot the normalized density $\rho_0/\rho_\infty$ of the dissipative vortex for $\nu=1$ (solid curve) and for $\nu=2$ (dashed curve). The inset of Fig.~\ref{dens_diss_final} exhibits the relative difference $\left(\rho_0-\rho_0^{\mathrm{nd}}\right)/\rho_0^{\mathrm{nd}}$
between the density $\rho_0$ of the dissipative vortex and the density $\rho_0^{\mathrm{nd}}$ of the standard non-dissipative vortex. Since the dissipation profile   
\eqref{gaussd} is concentrated around $\eta=0$, the difference between the dissipative and the non-dissipative solutions becomes negligible as one moves away from the center of the vortex. 
In Fig.~\ref{vel_diss_final} we plot the normalized radial velocity $|v_s|/c_{\infty}$ associated with the dissipative vortex for $\nu=1$ (solid curve) and for $\nu=2$ (dashed curve). We highlight that this velocity is strongly dependent on the corresponding dissipation profile and vanishes in the case of the standard non-dissipative vortex.

\section{\label{diss:super}Superradiance}
We now analyze the scattering of perturbations around a dissipative vortex and investigate the possibility of superradiance in such a setting. 
The first step towards this goal is to solve Eq.~\eqref{pertspinI}. We look for stationary solutions
\begin{equation}\label{chanvarI}
W_{\omega\lambda}(t,\bm{r})=e^{-i\left(\omega t-\lambda\varphi\right)}\widetilde{W}_{\omega\lambda}(s),  
\end{equation}
where
\begin{equation}\label{chanvarII}
\widetilde{W}_{\omega\lambda}(s)=\left[\begin{array}{c}\alpha(s)\\\beta(s)\end{array}\right]
\end{equation}
is defined in terms of the radius-dependent functions $\alpha(s)$ and $\beta(s)$. The parameter $\omega\in\mathbb{R}_+^*$ is the frequency of the excitation and the parameter $\lambda\in\mathbb{Z}$ is the azimuthal number of the excitation (not to be confused with the azimuthal number $\nu\in\mathbb{Z}$ of the background vortex solution).

The ansatz \eqref{chanvarI} transforms Eq.~\eqref{pertspinI} into 
\begin{equation}\label{numpertI}
\hbar\omega\widetilde{W}_{\omega\lambda}(s)=\left(\hat{\mathcal{L}}_{\nu,\lambda}+\hat{\mathcal{L}}_{\Gamma}\right)\widetilde{W}_{\omega\lambda}(s),\end{equation}
where $\hat{\mathcal{L}}_{\nu,\lambda}$ is a differential operator that encodes the non-dissipative features of the system, while $\hat{\mathcal{L}}_{\Gamma}$ is a differential operator that arises due to dissipation effects. Explicitly, $\hat{\mathcal{L}}_{\nu,\lambda}$ is the same operator studied in Ref.~\cite{carusottoI}, given by
\begin{equation}\label{numpertII}
    \hat{\mathcal{L}}_{\nu,\lambda}=\left[\begin{array}{cc}
        \displaystyle\mathcal{D}_{+} & \displaystyle U_0\rho_0(s) \\
        \displaystyle-U_0\rho_0(s) & \displaystyle\mathcal{D}_{-}
    \end{array}\right]
\end{equation}
with
\begin{equation}\label{numpertIIa}
    \mathcal{D}_{\pm}\equiv\pm\frac{\hbar^2}{2m}\left(-\frac{d^2}{ds^2}-\frac{1}{s}\frac{d}{ds} +\frac{(\nu\pm\lambda)^2}{s^2}\right)\pm2U_0\rho_0(s)\mp\mu .
\end{equation}
The operator $\hat{\mathcal{L}}_{\Gamma}$, on the other hand, is given by
\begin{align}\label{numpertIII}
    \hat{\mathcal{L}}_{\Gamma}=\frac{i\hbar}{2}\bigg[ &\Gamma(s) -2 v_s(s)\frac{d}{ds}-\bigg(\frac{v_s(s)}{s}+\frac{dv_s}{ds}(s)\bigg)\bigg]\mathbb{I}\notag\\+&\frac{mv_s(s)^2}{2}\sigma_z,
\end{align}
and includes both the dissipation parameter $\Gamma(s)$ and the non-zero radial velocity $v_s(s)$. Note that the off-diagonal terms in \eqref{numpertII} couple the two components of $\widetilde{W}_{\omega\lambda}(s)$. 

Eq.~\eqref{numpertI} can be solved analytically in two different asymptotic regimes. Near the origin ($s\rightarrow0$), the density of the vortex approaches zero, as seen in the specific cases represented in Fig.~\ref{dens_diss_final}. Consequently, near the center of the vortex, the two components of $\widetilde{W}_{\omega\lambda}(s)$ decouple from each other and Eq.~\eqref{numpertI} yields two independent equations for two unknowns. One can show that $s=0$ is a regular singular point of these decoupled equations if 
\begin{equation}
\lim\limits_{s \to0}s^2\Gamma(s)
\end{equation}
 exists, which is true for the Gaussian dissipation profile \eqref{gaussd}. Imposing as a boundary condition that 
the perturbation be finite at the origin, one finds that $\widetilde{W}_{\omega\lambda}(s)$ must be a linear combination of 
\begin{equation}\label{sols0}
\widetilde{W}_{\omega\lambda}^{0+}\approx \left[\begin{array}{c}s^{|\nu+\lambda|}\\0\end{array}\right] \text{ and } \, \widetilde{W}_{\omega\lambda}^{0-}\approx \left[\begin{array}{c}0\\ s^{|\nu-\lambda|}\end{array}\right]
\end{equation}
when $s\rightarrow 0$.

Secondly, far away from the vortex ($s\rightarrow\infty$), we have $\Gamma(s) \rightarrow 0$ and $\bm{v}_0(s)\rightarrow 0$, meaning that the condensed phase is approximately uniform, approaching a constant density $\rho_\infty$. The associated perturbations are caracterized by a constant speed $c_\infty$. 
In this regime, the four independent solutions of Eq.~\eqref{numpertI}, indexed by the parameter $j \in \{1,2,3,4\}$, are given by 
\begin{equation}\label{solsinfty}
\widetilde{W}_{\omega\lambda}^{k_j}(s)\approx\left[\begin{array}{c}\displaystyle1\\L_j\end{array}\right]\frac{e^{ik_js}}{\sqrt{s}},
\end{equation}
where $k_j$ are the associated wavenumbers that satisfy Bogolyubov's dispersion relation~\cite{BECbook}
\begin{equation}
\hbar^2\omega^2=\left(\hbar c_\infty k\right)^2+\left(\frac{\hbar^2k^2}{2m}\right)^2,
\end{equation}
and the parameters $L_j$ are given by
\begin{equation}
L_j = \frac{\hbar\omega-\left(\frac{\hbar^2k_j^2}{2m}+mc_\infty^2\right)}{mc_\infty^2}.
\end{equation}
Explicitly, the wavenumbers are
\begin{equation}\label{solsinftyk}
k_j=e^{\frac{ij\pi}{2}}\sqrt{2}\displaystyle{\frac{mc_\infty}{\hbar}}\sqrt{Q(\omega)+1},
\end{equation}
where 
\begin{equation}
Q(\omega) = \sqrt{1+\left(\frac{\hbar\omega}{\mu}\right)^2}.
\end{equation}
We note that the wavenumbers $k_1$, $k_2$, $k_3$, and $k_4$ correspond, respectively, to decaying, ingoing, growing and outgoing modes. 

In order to study scattering phenomena around the modified vortex, we define the in mode $\widetilde{W}^{in}_{\omega\lambda}(s)$ by requiring, in view of Eqs.~\eqref{sols0} and \eqref{solsinfty}, that
\begin{equation}\label{inmode}
\widetilde{W}^{in}_{\omega\lambda} \rightarrow \begin{cases} 
K_1 \widetilde{W}_{\omega\lambda}^{0+} + K_2 \widetilde{W}_{\omega\lambda}^{0-}, &s \rightarrow 0, \\ \\
A_{in}\widetilde{W}_{\omega\lambda}^{k_2}+A_R \widetilde{W}_{\omega\lambda}^{k_4}+E_d\widetilde{W}_{\omega\lambda}^{k_1}, &s \rightarrow \infty,
\end{cases}
\end{equation}
where $K_1$, $K_2$, $A_{in}$, $A_R$, $E_d$ are constants.
We also define the time-dependent in mode $W^{in}_{\omega\lambda}$, which is related to $\widetilde{W}^{in}_{\omega\lambda}$ through Eq.~\eqref{chanvarI}. This in mode represents an incoming wave with amplitude $A_{in}$ and frequency $\omega$ that scatters around the background vortex and is reflected back with amplitude $A_R$. Note that a decaying mode with amplitude $E_d$ is necessary for consistency. Additionally, the growing mode is discarded since we require the solution to be finite when $s\rightarrow\infty$. 

Any solution of Eq.~\eqref{pertspinI} for the perturbations must also satisfy the conservation equation \eqref{disscontspin}. Substituting the in mode $W^{in}_{\omega\lambda}$ into Eq.~\eqref{disscontspin} and integrating both sides over all space, one arrives at the following expression for the reflection coefficient $R$:
\begin{equation}\label{superI}
R=\left|\frac{A_R}{A_{in}}\right|^2=1+\frac{1}{|A_{in}|^2}I(\omega),
\end{equation}
where $I(\omega)$ is a function that encodes the effects of dissipation. It can be written as
\begin{equation}\label{superII}
I(\omega)=\frac{1}{4\pi\sqrt{2}c_\infty G(\omega)}\int_0^\infty\Gamma(s) \widetilde{ W}^{in\dagger}_{\omega\lambda}\sigma_z \widetilde {W}^{in}_{\omega\lambda} \,s \, ds,
\end{equation}
where 
\begin{equation}\label{superIII}
G(\omega) = \left[Q(\omega)^2-\frac{\hbar\omega}{\mu}Q(\omega)\right]\sqrt{Q(\omega)-1}.
\end{equation}

It is evident that, in the absence of dissipation, we have $I(\omega)=0$ and $R=1$: everything that is sent towards the vortex is reflected back to infinity. On the other hand, if $\Gamma$ is not zero, a net flux of energy appears. The direction of this net flux depends on the sign of $I(\omega)$. If $I(\omega)$ is positive, Eq.~\eqref{superI} implies that $R > 1$. In other words, the amplitude of the reflected wave is larger than the amplitude of the incident wave, caracterizing superradiant scattering. In such a case, the associated in mode $W^{in}_{\omega\lambda}$ is a superradiant mode that extracts energy from the vortex. 

Let us now understand what are the conditions that allow $I(\omega)$ to be positive. First, note that $G(\omega)$, defined by Eq.~\eqref{superIII}, is always positive. Additionally, remember that we have assumed in Sec.~\ref{theBEC:vortex} that $\Gamma(s) < 0$. Therefore, in view of Eq.~\eqref{superII}, we conclude that $I(\omega)$ can only be positive if the ``charge" density $\widetilde{W}^{in\dagger}_{\omega\lambda} \sigma_z \widetilde{W}^{in}_{\omega\lambda}$ is negative in a significant portion of the region where $\Gamma$ is non-negligible. Since we have assumed that the presence of a non-zero $\Gamma(s)$ is only relevant around the origin, we now look at the behaviour of the ``charge" density at the centre of the vortex. Using Eq.~\eqref{inmode}, we find that 
\begin{equation} \label{neg_charge}
\widetilde{ W}^{in\dagger}_{\omega\lambda}\sigma_z \widetilde {W}^{in}_{\omega\lambda}\rightarrow |K_1|^2s^{2|\nu+\lambda|} - |K_2|^2s^{2|\nu-\lambda|}
\end{equation} 
when $s \rightarrow 0$. When $\nu > 0$, assuming that both $K_1$ and $K_2$ are nonzero, we see that the contribution above will be dominated by the negative term if $\lambda > 0$. In other words, the ``charge" density at the centre of the vortex is negative if the wave and the vortex are corotating. When this happens, therefore, we expect superradiant scattering to occur. In particular, when $\nu=\lambda > 0$ and $s \rightarrow 0$, the righthand side of Eq.~\eqref{neg_charge} approaches a negative constant. Hence, we expect the energy extraction process to be more efficient when $\nu=\lambda$ than when $\nu \neq \lambda$.  

The superradiant character of a given mode depends on the specific values of the constants $K_1$ and $K_2$. Together with $A_R$ and $E_d$, those should be seen as functions of $A_{in}$, $\omega$ and all the other parameters in the problem. Unfortunately, they can only be determined numerically by solving the differential equation for the perturbations with appropriate boundary conditions. In other words, we need to first determine the in mode $\widetilde{W}^{in}_{\omega\lambda}$ numerically and then extract the associated reflection coefficient $R$. 

\subsection{Numerical Example}
The theoretical analysis presented above is valid for any dissipation profile $\Gamma$ that is concentrated around the origin. To illustrate the ocurrence of superradiance in a dissipative BEC vortex, we determine numerically the reflection coefficients associated with modes that scatter off the dissipation profile defined in Sec.~\ref{theBEC:vortex}. The background vortices we consider in our numerical analysis are, thus, the $\nu=1$ and the $\nu=2$ solutions of Eqs.~\eqref{addissMaGPEI} and \eqref{addissMaGPEII} shown in Figs.~\ref{dens_diss_final} and \ref{vel_diss_final}. We focus our investigation on the modes whose azimuthal number satisfy $|\lambda|\le \nu$. The details concerning our numerical implementation of the scattering problem is explained in the Appendix.

In Fig.~\ref{R_l=1} we display the  reflection coefficient $R$ associated with superradiant and non-superradiant modes that scatter off the $\nu=1$ vortex. The nonrotating ($\lambda = 0$) and the counterrotating ($\lambda=-1$) modes are characterized by $R<1$ regardless of their frequencies, meaning that these waves are always absorbed by the vortex. The reflection coefficient for corotating modes ($\lambda=1$), on the other hand, is greater than one for sufficiently low frequencies, demonstrating the ocurrence of superradiance. The regime of superradiant amplification is highlighted in the inset of Fig.~\ref{R_l=1}.

\begin{figure}[!htbp]
\centering
  \includegraphics[width = 0.99 \linewidth]{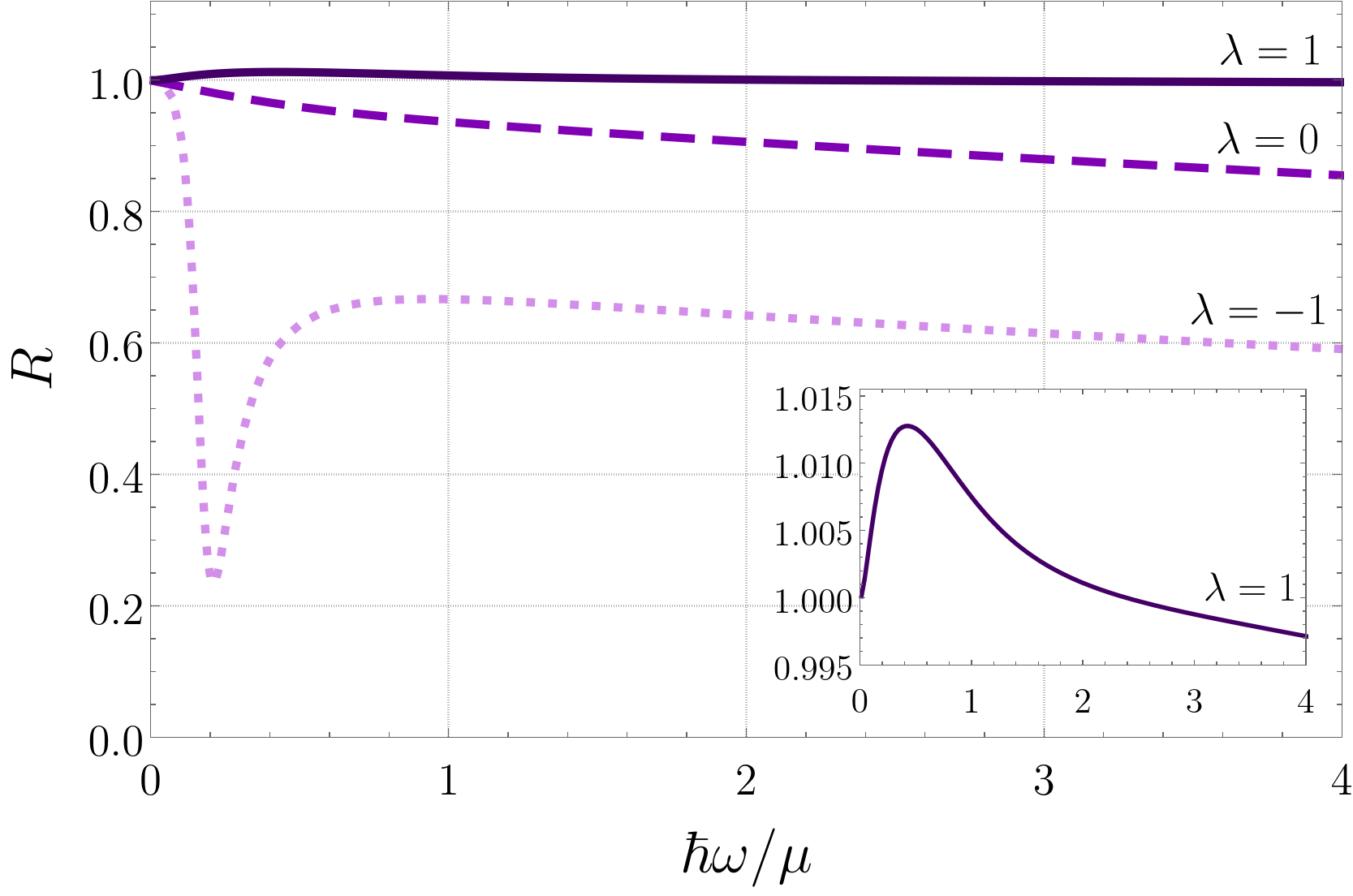}
  \caption{Reflection coefficient $R$ as a function of the frequency $\omega$ for  modes impinging on the dissipative quantum vortex $\nu=1$. The solid, dashed and dotted curves correspond, respectively, to $\lambda=1$, $\lambda=0$ and $\lambda=-1$. The physical parameters and the dissipation profile that determine the vortex are given in Sec.~\ref{theBEC:vortex}. The inset reveals the superradiant regime associated with the $\lambda=1$ mode.}
    \label{R_l=1}   
\end{figure}

\begin{figure}[!htbp]
\centering
  \includegraphics[width = 0.99 \linewidth]{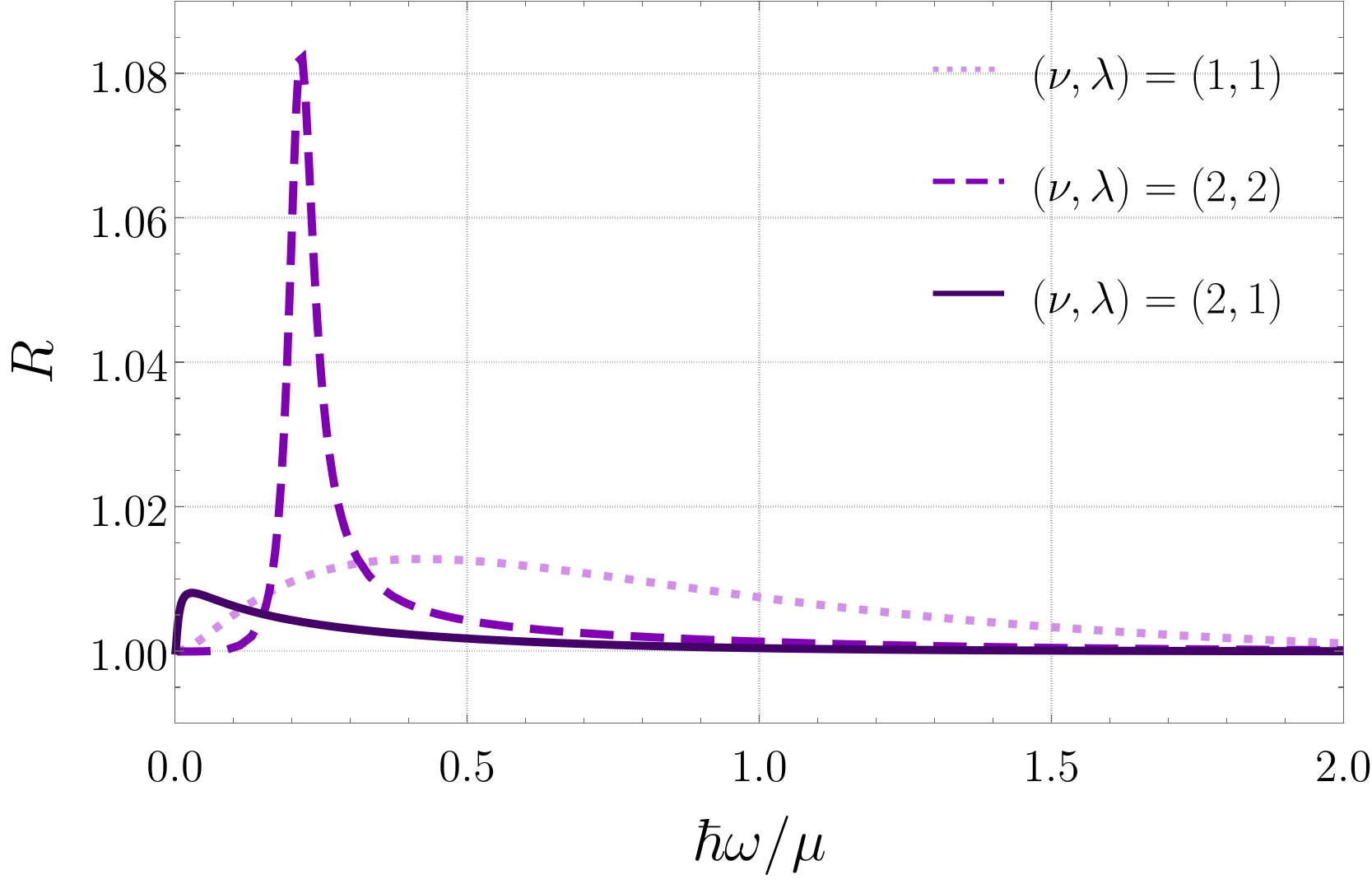}
  \caption{Reflection coefficient $R$ as a function of the frequency $\omega$ for superradiant modes impinging on the $\nu=1$ and the $\nu=2$ vortices. The dotted curve corresponds to the $\lambda=1$ mode of the $\nu=1$ vortex. The solid and dashed curves correspond, respectively to the $\lambda=1$ and the $\lambda=2$ modes of the $\nu=2$ vortex. The physical parameters and the dissipation profile that determine the vortices are given in Sec.~\ref{theBEC:vortex}.}
    \label{fig:superradiantmodes}   
\end{figure}

In Fig.~\ref{fig:superradiantmodes}, we exhibit the dependence of the reflection coefficient $R$ on the wave frequency $\omega$ for three different superradiant modes. The results allow the comparison of the superradiant regimes for the $0<\lambda\le \nu $ modes of the $\nu=1$ and the $\nu=2$ vortices. For each curve in Fig.~\ref{fig:superradiantmodes}, we compute the peak value $R_{\mathrm{p}}$ of the reflection coefficient and the associated peak frequency $\omega_{\mathrm{p}}$. We also determine the cutoff frequency $\omega_{\mathrm{c}}$ above which superradiance shuts off. The results for $R_{\mathrm{p}}$, $\omega_{\mathrm{p}}$ and $\omega_{\mathrm{c}}$  are displayed in Table \ref{Tab:cutoff} for each pair $(\nu,\lambda)$. In particular, we note that the maximum amplification observed is approximately $8\%$, associated with the $\lambda=2$ mode that scatters off the $\nu=2$ vortex.     
Compared to the other two curves in Fig.~\ref{fig:superradiantmodes}, we see that the superradiant peak is much sharper for the $(\nu,\lambda)=(2,2)$ curve. A similar resonance-like peak has been pointed out before in a different setting \cite{faccio} and might be related to the well-known fact that non-dissipative quantized vortices with $\nu>1$ are unstable, decaying into vortices with smaller winding numbers~\cite{carusottoI,patrick2021quantum,patrick2021origin}. Nonetheless, a more detailed study must be performed to investigate this possible connection.

\begin{table}[!htbp]
\begin{ruledtabular}
\begin{tabular}{llll}
$(\nu,\lambda)$ & $R_p$ & $\hbar\omega_p/{\mu}$ & $\hbar\omega_c/{\mu}$  \vspace{0.1cm} \\  
\hline
 \vspace{-0.3cm}
 \\
 (1,1) & 1.013 & 0.422 & 2.397 \\
(2,1) & 1.008 & 0.028 & 2.873 \\
(2,2) & 1.083 & 0.215 & 7.297  \\
\end{tabular}
\end{ruledtabular}
\caption{The maximum reflection coefficient $R_{\mathrm{p}}$, the associated peak frequency $\omega_{\mathrm{p}}$, and the cutoff frequency $\omega_{\mathrm{c}}$ for the superradiant modes exhibited in Fig.~\ref{fig:superradiantmodes}.}
\label{Tab:cutoff}
\end{table}

\section{\label{concl}Conclusion}

In this work, motivated by previous research on superradiance, we have studied the scattering of perturbations around a dissipative BEC vortex. More precisely, inspired by atom laser models, we have indroduced dissipation through an imaginary term in the Hamiltonian that describes the BEC. By solving the associated Gross-Pitaevskii equation, we have determined how the standard quantum vortex configurations are affected by dissipation. We have then investigated the prospect of superradiant amplification by these modified quantum vortices. After establishing the theoretical basis that support the phenomenon, we have provided a numerical study that illustrates the occurrence of superradiance.
The present analysis stands in parallel to other works that study superradiant phenomena around BEC configurations~\cite{superrad_bec22,superrad_bec3,superrad_bec4,superrad_bec5,superrad_bec6,
sam_dispersive,ergo4,carusottoI,carusottoII,carusottoIII,patrick2021quantum,patrick2021origin}. Here, a dissipation mechanism was necessary to take negative-norm modes out of the system and, hence, produce superradiance.

In particular, we have found stationary superradiant modes associated with the dissipative quantum vortices we considered. In contrast, in configurations without any dissipation mechanism~\cite{faccio,carusottoI}, it is suggested that superradiance still takes place as a transient effect. In such cases, as the background might be unstable, one has to keep track of the instability timescales when trying to measure superradiant effects. 
Therefore, a possible extension of our work would be to analyze the stability of the dissipative quantum vortex using the ideas presented in Ref.~\cite{carusottoI}. Another possibility in a similar direction would be to study quasinormal modes~\cite{Berti:2009kk,Konoplya:2011qq,Assumpcao:2018bka,Torres:2019sbr,Torres:2020tzs, Siqueira:2022tbc} around the dissipative quantum vortex. To accomplish this, one would have to modify the boundary conditions \eqref{inmode} to eliminate the incoming wave from infinity and solve the associated eigenvalue problem.

We emphasize that our work treats dissipation as something external to the background vortex and its excitations. Even though atom lasers from vortex lattices have been investigated previously~\cite{vortexatomlaser}, the ideas presented here can be a starting point for scenarios without external dissipation. In a real experiment, there are interactions between the vortex and the waves. From the point of view of the perturbations alone, the vortex itself acts as a sink (source), being able to absorb (emit) phonons and, by its own term, decay or be excited to a new state. In this sense, our work is also applicable if the interaction between vortex and waves can be modelled in terms of a dissipation profile $\Gamma(s)$. The main difference is that one would have to solve for both the perturbation and the background at the same time, i.e.~solve both Eqs.~\eqref{GPEloss} and \eqref{pertspinI} simultaneously. This article, together with Refs.~\cite{faccio,carusottoI}, suggests that superradiance should still take place in such a case.

Finally, we highlight the possibility of extending our results to other systems. Even though the present work was developed for cold atoms, Bose-Einstein condensation is a more general phenomenon that also occurs in other contexts. An interesting example, with applications in analogue gravity, is that of microcavity polaritons~\cite{polaritons1,polaritons2,jacquet2020polariton}.  These are quasi-particles of light and excitons (electron-hole pairs) that can be generated in suitably tailored semiconductors. The polaritons can be made to interact in a way very similar to particles in a dilute cold gas. In fact, their governing equation ressembles the Gross-Pitaevskii equation \eqref{GPEloss} if we associate the loss term with the finite lifetime of a polariton~\cite{jacquet2020polariton,lagoudakis2008quantized}. In particular, quantized vortices are possible~\cite{lagoudakis2008quantized,boulier2016injection,alperin2021multiply}. Such close connection with the present work suggests that the ideas discussed here should also find application in microcavity polariton fluids.

\section*{\label{ack}Acknowledgements}

The authors are grateful to the University of Nottingham for hospitality
while the initial stages of this work were developed. The authors would like to thank Sebastian Erne and Silke Weinfurtner for enlightening discussions.
This research was partially financed by the Coordena\c{c}\~ao de Aperfei\c{c}oamento de Pessoal de N\'ivel Superior (CAPES, Brazil) - Finance Code 001. The authors acknowledge financial support from the S\~{a}o Paulo Research Foundation (FAPESP, Brazil), Grants No.~2018/00048-7 and No.~2019/14476-3. M.~R.~also acknowledges support from the Conselho Nacional de Desenvolvimento Cient\'ifico e Tecnol\'ogico (CNPq, Brazil),
Grant No.~FA 315664/2020-7.

\appendix*

\section*{\label{apex}Appendix: Numerical Methods}

In this Appendix we go through the main ideas behind the numerical work of the present article. The whole procedure can be divided into two steps: first solving for the modified (dissipative) background configuration of the condensed phase of the BEC and then solving for its elementary excitations. Our goal is to find stationary solutions to the associated time-independent differential equations. In particular, we investigate the possibility of superradiant scattering around quantized vortices in the frequency domain following the general idea typically applied to other systems~\cite{superreview}. Nevertheless, one could also study superradiance by determining the time evolution of an initial state (e.g.~using a Fourier split operator method, as in Ref.~\cite{patrick2021quantum}, or the truncated Wigner approximation~\cite{sinatra}).    

\subsection*{\label{apex:I}Background Solution}
We use Chebyshev pseudo-spectral methods \cite{shapiro,boyd} to find the dissipative background solution. Explicitly, we consider Eqs.~\eqref{addissMaGPEI} and \eqref{addissMaGPEII} to determine quantum vortex configurations in the presence of dissipation. 
We first perform the change of variable $x=2\eta/(1+\eta)-1$ in order to bring the semi-infinite interval $[0,\infty)$ into the finite interval $[-1,1]$, where the Chebyshev polynomials are defined. Eqs.~\eqref{addissMaGPEI} and \eqref{addissMaGPEII}, in terms of the new variable $x$, become
\begin{align}\label{xI}
   \frac{(1-x)^3}{4(1+x)}&\frac{d}{dx}\left[\left(1-x^2\right)\frac{du}{dx}\right]\notag\\
   +&\left[1-\nu^2\left(\frac{1-x}{1+x}\right)^2\right]u-u^3-\frac{1}{2}uv^2=0
\end{align}
and
\begin{align}\label{xII}
\frac{1}{2}\frac{(1-x)^3}{1+x}\frac{d}{dx}\left[\left(\frac{1+x}{1-x}\right)u^2v\right]=u^2\widetilde{\Gamma},
\end{align}
where $u(x)=F(\eta(x))$ and $v(x)=\widetilde{v}_s(\eta(x))$. The associated boundary conditions for $u$ and $v$, taking into account Eqs.~\eqref{bcc1} and \eqref{bcc2}, become
\begin{equation} \label{bcc3}
 u(1)=1, \quad v(1)=0, \quad v(-1)=0. 
 \end{equation}

We expand the unknown functions $u$ and $v$ in terms of a basis of Chebyshev polynomials $\{T_0(x),T_1(x),...,T_N(x)\}$:
\begin{equation}\label{expChebI}
    u(x)=\displaystyle\sum_{i=0}^NU_iT_i(x)
\end{equation}
and
\begin{equation}\label{expChebII}
    v(x)=\displaystyle\sum_{i=0}^NV_iT_i(x).
\end{equation}
Ideally the expansions above would contain infinite terms, but truncation at some finite number $N$ is necessary for numerical purposes. We thus have $2(N+1)$ unknown coefficients $\{U_0,U_1,...,U_N,V_0,V_1,...,V_N\}$.
We set a grid $x_j$ on our domain based on the extrema of the highest-order polynomial $T_{N}(x)$~\cite{shapiro}:
\begin{equation}\label{extr}
    x_j=\cos\left(j\frac{\pi}{N}\right),\text{ }j=0,1,2,...,N.
\end{equation}
Since $x_0=1$ and $x_N=-1$ are the boundary points of our domain, Eqs.~\eqref{bcc3},  \eqref{expChebI} and \eqref{expChebII} imply that
\begin{equation}\label{bc}
     \sum_{i=0}^NU_i = 1 , \quad
     \sum_{i=0}^NV_i = 0 , \quad
     \sum_{i=0}^NV_i(-1)^i =0.
\end{equation}

We also substitute expansions \eqref{expChebI} and \eqref{expChebII} back into Eqs.~\eqref{xI} and \eqref{xII}, and apply the resulting expressions into the $2(N-1)$ inner points of the grid. This enforces that Eqs.~\eqref{xI} and \eqref{xII} hold exactly at these grid points, and produces $2(N-1)$ equations. The remaining equations are \eqref{bc} together with either \eqref{xI} or \eqref{xII} applied at $x=-1$.
The problem now consists of solving a non-linear system of $2(N+1)$ equations for the $2(N+1)$ expansion coefficients. This is carried out using the Newton-Raphson method. We set $N=50$ in our calculations and take $u(x)=1$, $v(x)=0$ as the initial guess for the solution.

\subsection*{\label{apex:II}Perturbations}

We use a second-order finite difference method to solve Eq.~\eqref{numpertI} for the perturbations $\widetilde{W}^{in}_{\omega\lambda}$. We start by setting up the (adimensionalized) domain $[\eta_{min},\eta_{max}]$, where $\eta_{min}$ is sufficiently close to the origin and $\eta_{max}$ is sufficiently far away from the vortex. In order to better capture the behavior of the system, we use a non-uniform grid $\eta_i$ consisting of two uniform subgrids joined at a common point $\eta_P$. In other words, subgrid 1 covers $[\eta_{min},\eta_P]$ with step $\Delta\eta_1$ and subgrid 2 covers $[\eta_P+\Delta\eta_2,\eta_{max}]$ with step $\Delta\eta_2$. Inside each subgrid we take the following second-order finite difference approximations for the derivatives:
\begin{equation}\label{FDI}
\left.\frac{d\widetilde{W}^{in}_{\omega\lambda}}{d\eta}\right|_{\eta_i} \! \! \! \approx\frac{\widetilde{W}^{in}_{\omega\lambda}(\eta_{i+1})-\widetilde{W}^{in}_{\omega\lambda}(\eta_{i-1})}{2\Delta\eta}
\end{equation}
and
\begin{equation}\label{FDII}
\left.\frac{d^2\widetilde{W}^{in}_{\omega\lambda}}{d\eta^2}\right|_{\eta_i} \! \! \! \approx\frac{\widetilde{W}^{in}_{\omega\lambda}(\eta_{i+1})-2\widetilde{W}^{in}_{\omega\lambda}(\eta_{i})+\widetilde{W}^{in}_{\omega\lambda}(\eta_{i-1})}{\Delta\eta^2},
\end{equation}
where
\begin{equation}\label{deltaeta}
\Delta \eta = \begin{cases}
\Delta \eta _1, \quad & \text{if \  }\eta_i<\eta_P,\\
\Delta \eta _2, \quad & \text{if  \ }\eta_i>\eta_P.
\end{cases}
\end{equation}
Exactly at $\eta_P$, however, the formulas are more involved due to the change of subgrids. The expressions we use for the finite difference approximations of the derivatives at $\eta_i=\eta_P$ can be found in Ref.~\cite{nonuniform}. 

The result of the discretization procedure is a linear system of equations for the values of the functions $\alpha(\eta)$ and $\beta(\eta)$ on each point of the non-uniform grid. We need to complement this system of equations with the boundary conditions discussed in Sec.~IV. In fact, the boundary conditions we need to implement are given by expression \eqref{inmode}. Without loss of generality we set $A_{in}=1$. Since the parameters $K_1$, $K_2$, $A_R$ and $E_d$ are not known, we need to manipulate expression \eqref{inmode} and its derivative to eliminate them. Additionally, taking into account the fact that $\eta_{min}\neq 0$ and $\eta_{max} \neq \infty$, we consider series expansions of the mode solutions around $\eta = 0$ and $\eta = \infty$ in order to improve accuracy. 

More precisely, including higher order terms and using the dimensionless variable $\eta$, expression \eqref{inmode} around the origin becomes
\begin{equation}\label{frobexp}
   \widetilde{W}^{in}_{\omega\lambda} =  \left[\begin{array}{c}
    \alpha(\eta)\\
    \beta(\eta)
\end{array}\right] = \left[\begin{array}{c}
    \eta^{|\nu+\lambda|}K_1\left(1+\sum a_i\eta^i \right)\\
    \eta^{|\nu-\lambda|}K_2\left(1+ \sum b_i\eta^i \right)
\end{array}\right], \vspace{0.2cm}
\end{equation}
where the constant coefficients $a_i$ and $b_i$ ($i \in \mathbb N^*$) can be calculated up to any desired order (in terms of $\omega$, $\lambda$, and the background parameters) by substituting \eqref{frobexp} into \eqref{numpertI}. The unknown constants $K_1$ and $K_2$ can be eliminated by manipulating the expression above and its derivative at $\eta=\eta_{min}$. We thus obtain the following Robin boundary condition: 
\begin{equation} \label{robin1}
\left. \frac{d \widetilde{W}_{\omega\lambda}^{in}}{d \eta} \right|_{\eta_{min}} \! \!   + M_{\omega\lambda} \widetilde{W}_{\omega\lambda}^{in}(\eta_{min}) = 0,
\end{equation}   
where $M_{\omega\lambda}$ is a known $2 \times 2$ matrix whose off-diagonal terms are of subleading order in powers of $\eta_{min}$.

On the other hand, including higher order terms in the expression for the Bogolyubov modes \eqref{solsinfty} and using the dimensionless variable $\eta$, one obtains (for each $j \in \{1,2,3,4\}$):
\begin{equation}\label{infexp}
   \widetilde{W}_{\omega\lambda}^{k_j}(\eta) =   \left[\begin{array}{c}
    \alpha(s)\\
    \beta(s)
\end{array}\right]\approx \left[\begin{array}{c}1 + \sum A_{ji}\,  \eta^{-i}\\L_j + \sum B_{ji} \, \eta^{-i} \end{array}\right] \frac{e^{ik_j \eta}}{\sqrt{\eta}},
\end{equation}
where the constant coefficients $A_{ji}$ and $B_{ji}$ ($i \in \mathbb N^*$) can be calculated up to any desired order (in terms of $\omega$, $\lambda$, and the background parameters) by substituting \eqref{infexp} into \eqref{numpertI}. We then substitute the expansions above for $\widetilde{W}_{\omega\lambda}^{k_j}$ into the $s \rightarrow \infty$ limit of expression $\eqref{inmode}$. We can eliminate the unknowns $A_R$ and $E_d$ by manipulating the asymptotic limit of $\widetilde{W}_{\omega\lambda}^{in}$ and its derivatives. As before, we obtain a mixed-type boundary condition at $\eta=\eta_{max}$:
\begin{equation} \label{robin2}
\left. \frac{d \widetilde{W}_{\omega\lambda}^{in}}{d \eta}\right|_{\eta_{max}} \! \! + N_{\omega\lambda} \widetilde{W}_{\omega\lambda}(\eta_{max}) = 0,
\end{equation}   
where $N_{\omega\lambda}$ is another known $2 \times 2$ matrix whose off-diagonal terms are of subleading order in powers of $\eta_{max}^{-1}$.

In the end, the linear system of equations obtained from the discretization of the differential equation \eqref{numpertI} and the boundary conditions \eqref{robin1}, \eqref{robin2} is solved through an LU decomposition method. Convergence tests were performed to determine the appropriate grid parameters and ensure the correcteness of the numerical results. We have set the grid parameters as $\eta_{min}=10^{-10}$, $\eta_{max}=5000$, $\eta_P=10$. The number of grid points used was $1000$ in subgrid 1 and $5000$ in subgrid 2. The only exception to this choice of parameters was the regime of small frequencies (i.e.~$\hbar \omega /\mu 	< 0.01$), for which we increased the end point $\eta_{max}$ to $100000$ and the number of points in subgrid 2 to $10^6$ (while keeping the other parameters unchanged).

\bibliography{references}

\end{document}